\documentclass[]{WileyMSP-template}
\usepackage{subfigure}
\usepackage{dcolumn}
\begin{document}

\pagestyle{fancy}

\title{Spatiotemporal Co-reflection with Spacetime Discontinuities at Moving Interfaces}

\maketitle


\author{Yongge Wang}
\author{Jingfeng Yao*}
\author{Chengxun Yuan$^\dagger$}
\author{Zhongxiang Zhou$^\ddagger$}



\begin{affiliations}
\\
Yongge Wang\\
School of Physics, Harbin Institute of Technology, Harbin 150001, People’s Republic of China\\
Jingfeng Yao$^*$, Chengxun Yuan$^\dagger$, Zhongxiang Zhou$^\ddagger$\\
School of Physics, Harbin Institute of Technology, Harbin 150001, People’s Republic of China\\
Heilongjiang Provincial Key Laboratory of Plasma Physics and Application Technology, Harbin 150001, People’s Republic of China\\
Email Address:*yaojf@hit.edu.cn, ${}^\dagger$yuancx@hit.edu.cn, ${}^\ddagger$zhouzx@hit.edu.cn
\end{affiliations}


\keywords{Spatiotemporal metamaterials, Spatiotemporal interface, Negative refraction}

\begin{abstract}\\
\textbf{Abstract}	
	\\
The control of reflection and refraction at interfaces using engineered media is central to numerous optical technologies, with negative refraction and the suppression of backscattering representing two prominent research frontiers. In this work, we demonstrate that an effective negative refraction accompanied by an absence of backscattering can be realized at a moving spatiotemporal interface when temporal and spatial reflections occur concurrently. While such spatiotemporal co-reflection is prohibited in one-dimensional linear dispersive media, we show that it becomes permissible under oblique incidence within a specific range of traveling-wave modulation velocities. Leveraging this mechanism, we propose a spatiotemporal flat lens capable of nonreciprocal electromagnetic wave focusing. These findings provide a framework for developing advanced spatiotemporal metamaterials and time-varying metasurfaces.
\\
\end{abstract}


The manipulation of electromagnetic waves at interfaces forms the physical cornerstone of modern optical technologies, ranging from sub-wavelength imaging and beam steering to signal processing and electromagnetic cloaking\cite{PhysRevLett.85.3966,doi:10.1126/science.1133628}. Among these phenomena, negative refraction has attracted significant attention due to its counter-intuitive ability to bend light toward the same side of the interface normal as the incident beam, enabling exotic effects such as perfect lensing and super-resolution\cite{Veselago:1968,doi:10.1126/science.1058847}. Traditionally, negative refraction is realized using artificial media with simultaneous negative permittivity and permeability, such as metamaterials and photonic crystals\cite{Soukoulis2011PastAA,PhysRevLett.92.127401}. Also, metasurfaces based on the generalized Snell’s law have provided an alternative route by introducing abrupt phase discontinuities\cite{doi:10.1126/science.1210713,Aieta2012AberrationfreeUF}. However, these spatial approaches typically rely on periodic sub-wavelength resonators that break the translational symmetry along the interface, leading to the non-conservation of the tangential wave-vector component.

An emerging paradigm in wave control involves the dynamic modulation of material properties in the time domain\cite{PhysRevLett.133.186902,PhysRevLett.133.083802,PhysRevLett.125.127403,PhysRevLett.133.263802}. When a wave experiences an abrupt temporal change in electromagnetic parameters, it undergoes scattering processes known as temporal reflection and refraction\cite{PhysRevA.79.053821,WOS:001604755900027,WOS:001287600900019,WOS:000531425900022}. Temporal reflection, in particular, results in the inversion of the wave propagation direction, acting as a physical realization of phase conjugation. Notable advances include the demonstration of topological edge states at temporal boundaries, wave amplification in photonic time crystals, and radiation from stationary charges\cite{Lustig:18,5hf5-pg3t,doi:10.1126/science.abo3324,li2023stationary}. These phenomena have been experimentally demonstrated across diverse platforms, including transmission lines, metasurfaces, water waves, and elastic systems\cite{10.1063/1.4928659,doi:10.1126/sciadv.adg7541,WOS:001287600900019,Bacot2015TimeRA,Wang2025ExperimentalRO}.

For a wave obliquely incident on a spatial interface, if temporal and spatial reflections occur simultaneously, both the frequency and the normal wave-vector component undergo a sign reversal. This dual inversion forces the scattered wave to propagate on the same side of the interface normal as the incident wave, mimicking the signature of negative refraction without double-negative media or breaking translational symmetry. While previous studies on abrupt interfaces have observed anomalous refraction\cite{PhysRevLett.132.263802}, time reflection and spatial reflection appear not to be truly simultaneous but rather two consecutive processes.

In this Letter, we propose a mechanism to realize simultaneous temporal and spatial reflection at a moving modulation interface, a process we term spatiotemporal co‑reflection. Spatiotemporal interfaces offer a degree of freedom inaccessible to purely spatial or purely temporal boundaries. Recently, scattering at moving modulated interfaces has been experimentally observed\cite{WOS:001604755900027}. While early investigations into moving interfaces found that either spatial or temporal reflection dominates exclusively\cite{10.1063/1.1655718,PhysRevB.107.115129,PhysRevApplied.20.054029,https://doi.org/10.1002/lpor.202300130}, we demonstrate that spatiotemporal co-reflection is fundamentally prohibited in one-dimensional linear dispersive media. By examining the conditions required for concurrent reflection, we show that this regime is physically accessible in  one-dimensional Drude media and two dimensional system. These findings are universal and can be extended to other wave physics systems, such as water waves and elastic waves.

\begin{figure}[htp]
	\centering
	\includegraphics[width=0.75\linewidth]{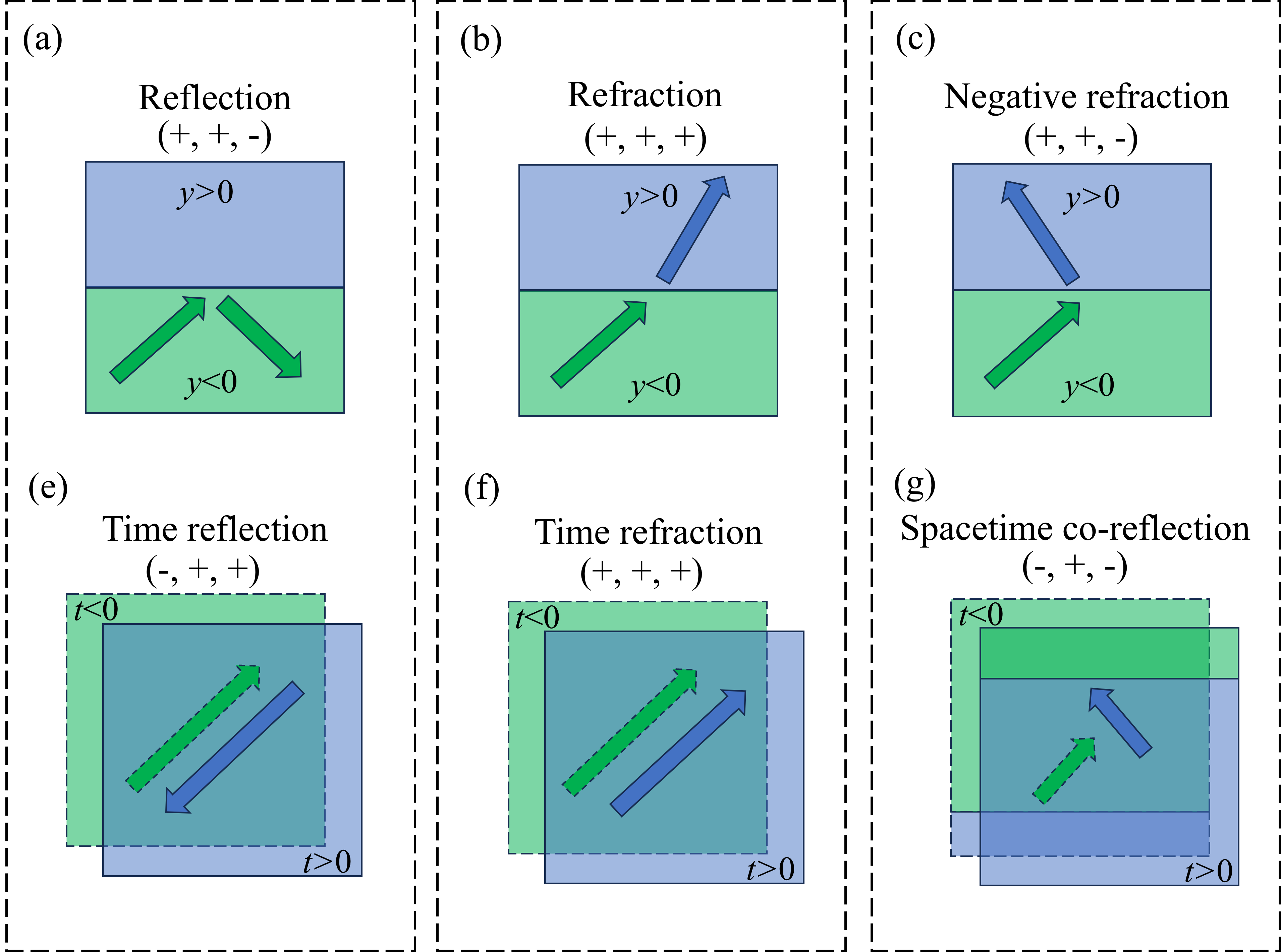}
	\caption{Several different phenomena of reflection and refraction, where the changes in the signs of the three parameters $(\omega,k_\parallel,k_\perp)$ after reflection/refraction are indicated. "+" denotes no sign change, and "–" denotes a sign change. (a) Spatial reflection; (b) spatial refraction; (c) negative refraction; (e) temporal reflection; (f) temporal refraction; (g) simultaneous temporal and spatial reflection at a moving interface. The two colors, blue and green, represent two media with different electromagnetic parameters. }
	\label{fig1}
\end{figure}

Here, we demonstrate several possible effects at spatiotemporal interfaces in Figure \ref{fig1}. Without loss of generality, we consider an incident electromagnetic wave propagating in the $x–y$ plane, with the interface oriented along the $x$ direction. The sign changes of the frequency and the wave-vector components parallel and perpendicular to the interface, $(\omega,k_\parallel,k_\perp)$, during the interfacial interaction are indicated in the figure. As depicted in Figure \ref{fig1}(a), spatial reflection is characterized by a sign reversal of the transverse wave-vector component $k_\perp$, whereas temporal refraction preserves the signs of both the frequency and the wave vector (Figure \ref{fig1}(b)). Notably, while negative refraction similarly entails a sign change of the normal wave vector component, meaning that the interfacial scattering process phenomenologically identical to spatial reflection. 

The essential difference originates from the material response: the wave vector is antiparallel to the Poynting vector, so that energy flows away from the interface, giving rise to negative refraction. Negative refraction necessitates simultaneous negative permeability and permittivity, a condition conventionally realized through artificial structures such as gratings, metamaterials, or photonic crystals. These media possess periodic spatial features that fundamentally break the translational symmetry along the interface. Consequently, the tangential wave-vector component is not conserved.

In contrast to spatial reflection, both time-reflected and time-refracted waves propagate into the new medium ($t>0$), as illustrated in Figure \ref{fig1}(e) and 1(f). These two processes differ in their fundamental behavior: temporal refraction preserves the signs of both the frequency and the wave vector, while temporal reflection is characterized by a sign reversal of the frequency. This frequency inversion is physically equivalent to the process of phase conjugation. However, when an interface supports simultaneous temporal and spatial reflections, both the frequency and the normal wave-vector component undergo a concurrent sign reversal, as shown in Figure \ref{fig1}(g). Under these conditions, the space-time co-reflected wave propagates on the same side of the interface normal as the incident wave, effectively realizing a regime of effective negative refraction.

\begin{figure*}[tp]
	\centering
	\includegraphics[width=0.85\linewidth]{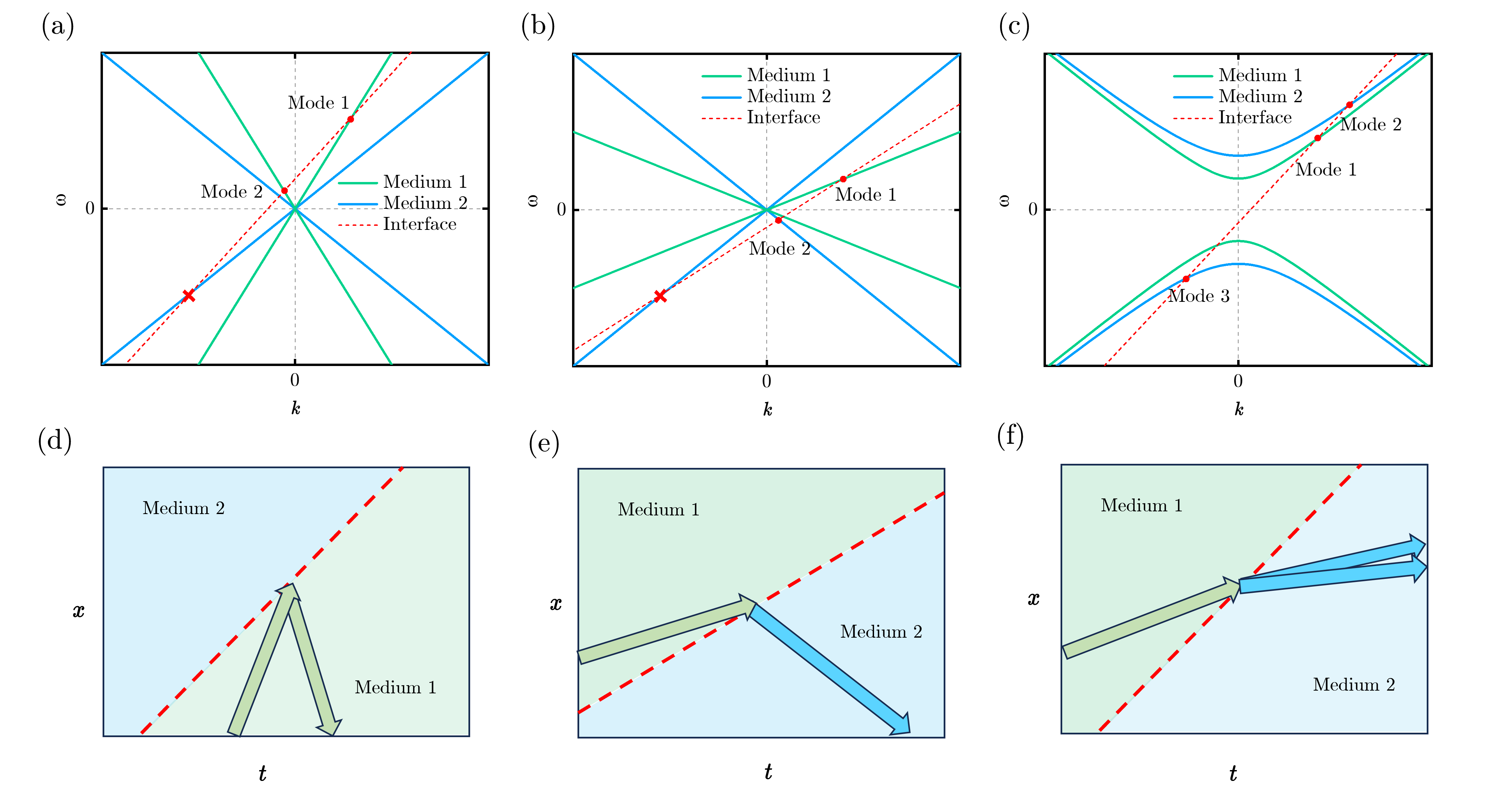}
	\caption{Schematic of scattering behaviors of a wave at spatiotemporal interfaces. Here, mode 1 is the initial mode. (a–c) Mode conversion induced by a moving interface in dispersive media, plotted on dispersion relation diagrams. (d–f) Spatiotemporal evolution of the scattering of an incident wave at a moving interface. (a) and (d) correspond to linear dispersion with the initial pulse in the lower‑refractive‑index medium. (b) and (e) correspond linear dispersion with the initial pulse in the higher‑refractive‑index medium. (c) and (f) are in Drude dispersive medium. }
	\label{fig2}
\end{figure*}

A natural approach involves considering a moving modulation profile where the permittivity is defined by $\varepsilon(y-vt)$, with $v$ representing the interface velocity. As $v$ increases from zero to infinity, the wave-interface interaction undergoes a transition from the spatial scattering regime to the temporal regime. Within an intermediate velocity range, spatial and temporal reflections could potentially occur simultaneously, manifesting as an effective negative refraction. However, we demonstrate that this spatiotemporal co-reflection is fundamentally unattainable in one dimensional media with linear dispersion under such moving modulation. Consequently, this specific scattering mode has remained largely unexplored in previous literature\cite{wang2026breakinglimitationstemporalmodulation}.

Upon interaction with a moving interface, spatiotemporal translational invariance dictates that $\omega-kv$ remains a conserved quantity of the system, where $v$ denotes the interface velocity. Consequently, the incident and scattered waves must satisfy the following invariant:
\begin{equation}
	\omega_+-vk_+=	\omega_--vk_-
\end{equation}
Here, $(\omega_-,k_-)$ and $(\omega_+,k_+)$ represent the frequency and wave-vector components of the incident and scattered waves, respectively. In addition to this matching condition, the scattered modes must satisfy the constitutive dispersion relation of the medium. The permitted scattering channels at the spatiotemporal interface can thus be identified graphically within the band diagram, as shown in Figure \ref{fig2}.

Figure \ref{fig2}(a) and \ref{fig2}(b) correspond to the case of linear dispersive media with refractive indices $n_1$ and $n_2$. When the interface velocity $v$ is superluminal relative to both media, $v>\max(c/n_1,c/n_2)$, the scattering process is strictly time-like, meaning that reflection is accompanied by a sign reversal of the frequency. Here, $c$ is the speed of light. Conversely, for the subluminal case where $v<\min(c/n_1,c/n_2)$, the scattering is space-like, characterized by a reversal of the wave vector. These two conventional regimes are not depicted in Figure \ref{fig2}.

When the interface velocity lies between the phase velocities of the two media, two distinct total reflection phenomena emerge. As illustrated in Figure \ref{fig2}(a), if the incident mode (Mode 1) is situated in the medium with faster wave speed, the wave overtakes the interface and undergoes scattering. Mode 2 represents a space-like reflection returning to medium 1. Although the phase-matching line intersects the dispersion relation of medium 2 in the third quadrant, this mode is forbidden because its wave speed is lower than the interface velocity, precluding it from entering medium 2. Consequently, as shown in Figure \ref{fig2}(d), only a single space-like reflected mode exists.

In Figure \ref{fig2}(b), the incident mode originates in the slower medium, allowing the moving interface to overtake the wave. A time-reflected wave appears in medium 2, corresponding to the intersection of the phase-matching line and the dispersion curve in the fourth quadrant. While an intersection also exists in the third quadrant for medium 2, it is similarly precluded because the wave speed there exceeds the interface velocity. Thus, as depicted in Figure \ref{fig2}(e), only a single time-reflected wave is supported. We therefore conclude that spatiotemporal co-reflection cannot be simultaneously realized at a single moving interface within one-dimensional linear dispersive media.

In a Drude dispersive medium, however, this spatiotemporal co-reflection becomes physically realizable. As illustrated in Figure \ref{fig2}(c), assuming a medium characterized by Drude dispersion, the phase-matching line for an initial mode 1 intersects the dispersion curves of medium 2 in both the first and third quadrants. Since the interface velocity $v$ in this regime exceeds the group velocity within the medium, both modes are permitted to propagate into medium 2. Consequently, as shown in Figure \ref{fig2}(f), two distinct effective refractive waves are generated. Crucially, while one represents a standard spatiotemporal refracted wave (mode 2), the other (mode 3) is a spatiotemporal co-reflected wave characterized by a simultaneous sign reversal of both frequency and wave vector. This mechanism and the resulting phenomenology differ fundamentally from the behavior observed in linear dispersive media. It should be noted that we have excluded the zero-frequency modes of the Drude medium from Figure \ref{fig2}(c) and \ref{fig2}(f). The excitation of such modes depends on the continuity conditions inherent to temporal modulation, a topic addressed in prior literature.
\begin{figure}[htp]
	\centering
	\includegraphics[width=0.4\linewidth]{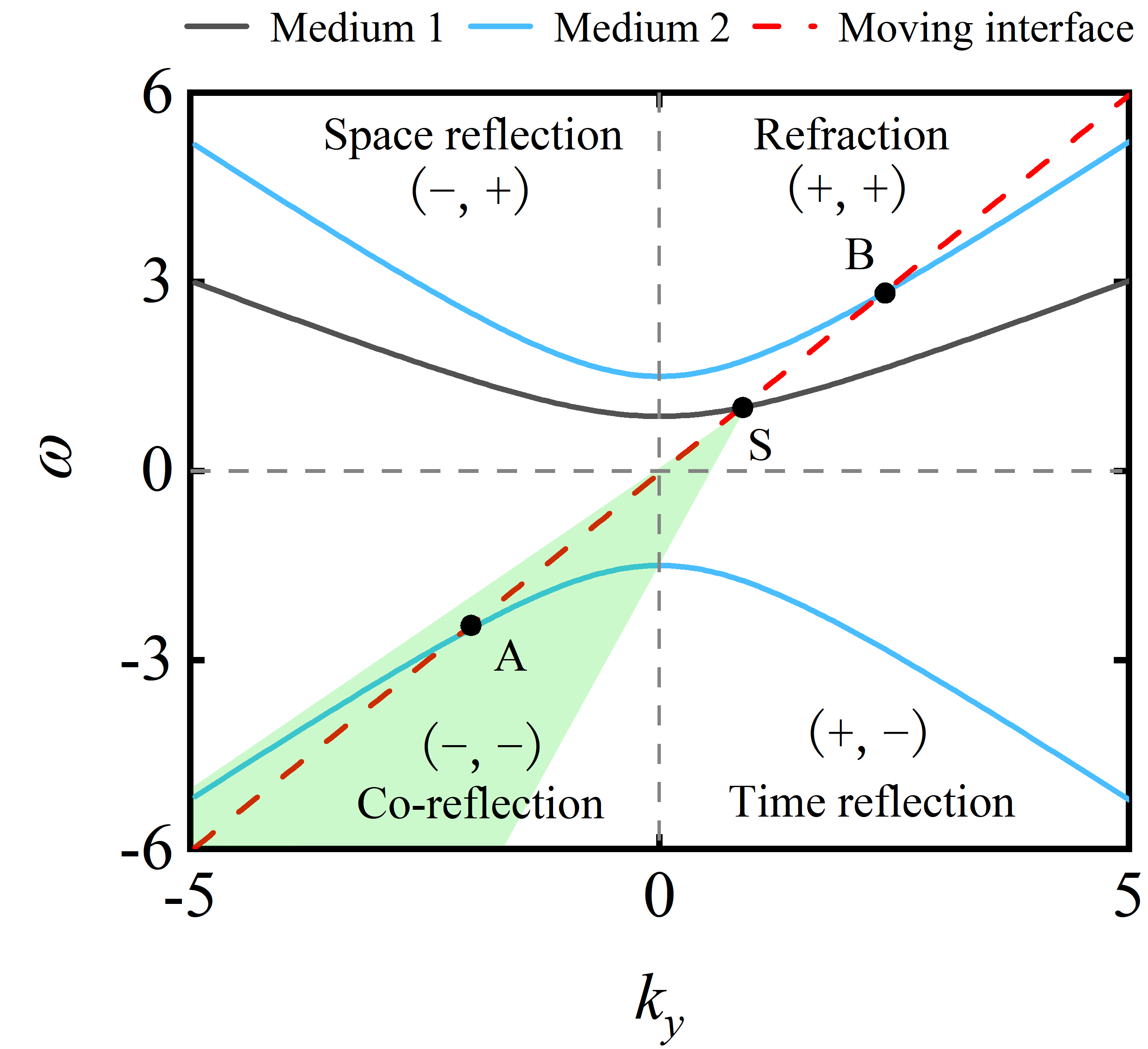}
	\caption{Schematic of spatiotemporal refraction and co-reflection under oblique incidence. The scattering phenomenology at a moving interface is illustratedwhere the respective sign changes of the frequency and normal wave-vector component, $(k_y,\omega)$, are labeled for each quadrant to characterize the interaction. Here, he initial mode is designated as S. The slope of the red dashed line represents the characteristic velocity $v$ of the moving interface. The shaded area indicates the modulation velocity window supporting spatiotemporal co-reflection.}
	\label{fig3}
\end{figure}

For the case of oblique incidence, we once again consider a linear medium, with the $(\omega,k_y)$ dispersion curves presented in Figure \ref{fig3}. Notably, these curves exhibit a hyperbolic profile similar to that observed in Figure \ref{fig2}(f). This structural similarity implies that spatiotemporal co-reflection at a moving interface is also permissible for linear media under oblique incidence.

We analyze the scattering of an initial mode S originating in medium 1. The resulting spatiotemporal scattering modes are mapped across the four quadrants, characterized by the respective sign changes in frequency and the normal wave-vector component. With the initial mode S fixed, spatiotemporal co-reflection is permitted when the slope of the phase-matching line falls within the shaded velocity window. In this superluminal modulation regime, both output modes act as effective refractive modes, with no equivalent reflective modes present (i.e., only a single type of reflection occurs). Physically, the spatiotemporal co-reflection mode corresponds to the Figure \ref{fig1}(g), representing an effective regime of negative refraction.

\begin{figure*}[htp]
	\centering
	\includegraphics[width=0.85\linewidth]{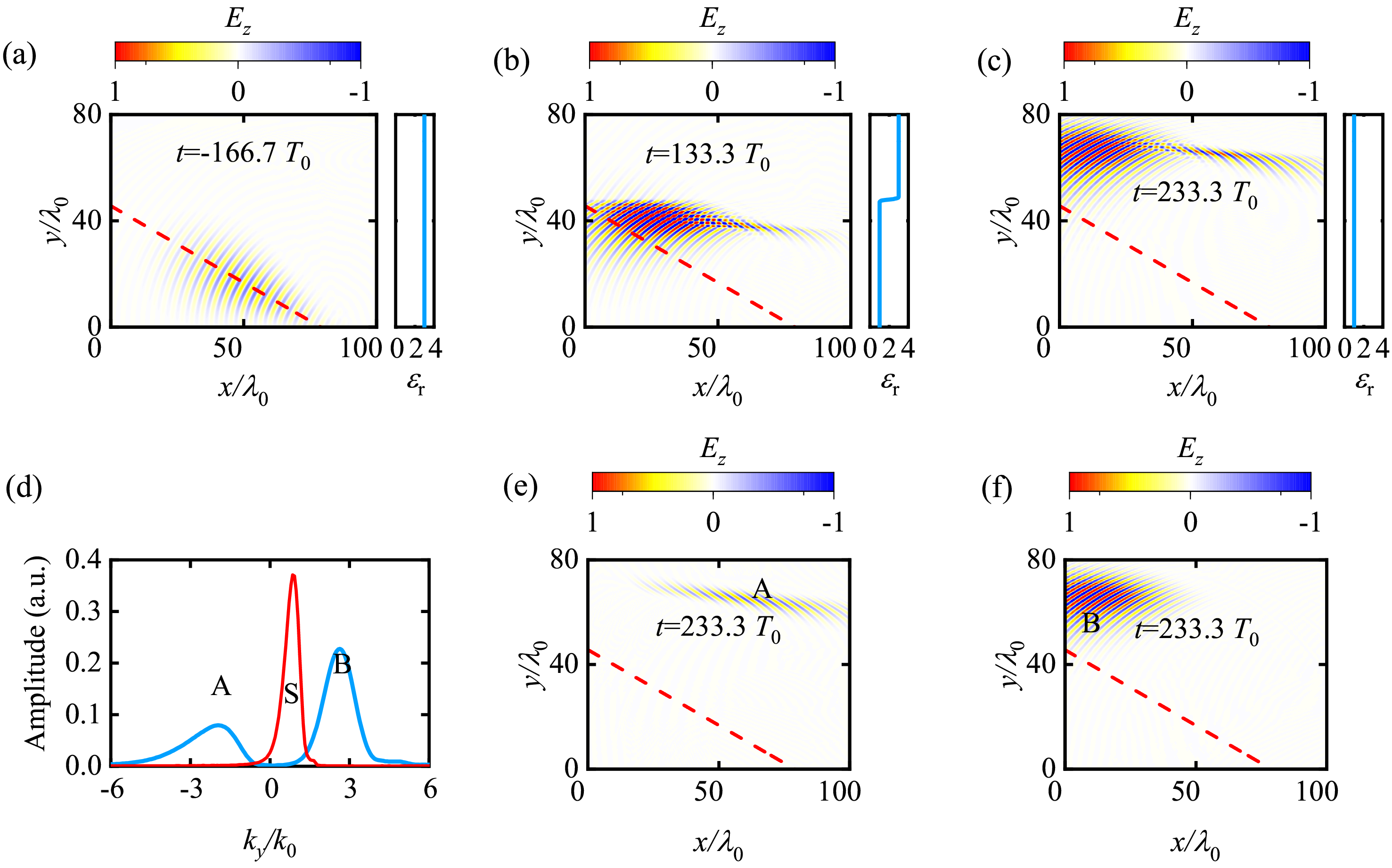}
	\caption{Scattering of an obliquely incident pulse at a moving modulation interface. The red dashed line indicates the propagation direction of the incident pulse. (a)–(c) Spatial evolution of the pulse at discrete time; the corresponding spatial distributions of the permittivity $\varepsilon$ are labeled to the right of each panel. (d) Calculated distribution of the incident pulse (A) and scattered pulses (A and B) in $k_y$-space. The two distinct peaks, labeled A and B in $k_y$-space, are isolated and plotted in (e) and (f), corresponding to the negative and positive refraction modes, respectively. }
	\label{fig4}
\end{figure*}

To validate the aforementioned theory, we perform Finite-Difference Time-Domain (FDTD) simulations of an obliquely incident wave interacting with a moving interface. 
The media are modeled with linear dispersion, and the continuity of the electric displacement field is enforced at the temporal discontinuity. The pulse center frequency is set to $\omega_0$, with a corresponding free-space wave vector $k_0=\omega_0/c$. The initial medium is assigned a permittivity of $\varepsilon_1=3$, and the incident pulse is characterized by a central transverse wave vector of $1.5k_0$. The second medium has a permittivity of $\varepsilon_2$. The interface moves at a velocity of $1.2\,c$, situating the system in the superluminal modulation regime. Under these configurations, spatiotemporal co-reflection is physically permitted. The resulting spatiotemporal evolution of the pulse is illustrated in Figure \ref{fig4}.

In Figure 4(a)–4(c), the red dashed line denotes the propagation direction of the incident pulse, with the spatial distribution of the permittivity $\varepsilon$ at corresponding time steps provided to the right of each field profile. Following the interaction with the moving interface, the initial pulse splits into two distinct components. Notably, both secondary pulses maintain forward propagation along the positive $y$-direction; consequently, the system exhibits no equivalent reflective behavior.

In our numerical study, the FDTD simulations are implemented using complex-valued fields. While the underlying physics remains consistent with real-valued formulations, this approach allows the determination of the wave-vector signs. Figure \ref{fig4}(d) presents the spectral distribution of the electric field transformed into the $k_y$-domain. The incident pulse, prior to scattering and corresponding to the spatial profile in Figure \ref{fig4}(a), is labeled as S. It appears as a single peak in the positive $k_y$ region.

Following the interaction with the moving interface, the spectral distribution, which corresponds to the spatial field in Figure \ref{fig4}(c), is shown by the blue curve in Figure \ref{fig4}(d). The spectrum clearly resolves into two distinct peaks situated on opposite sides of $k_y=0$. This spectrum split indicates that Mode A has undergone a spatial reflection, whereas Mode B has experienced only spatial refraction. To further elucidate the nature of these modes, we divided the negative-mode A and positive-mode B in $k_y$-space and performed an inverse transform back to the spatial domain, as shown in Figure \ref{fig4}(e) and (f).
A comparison with the real-space evolution in Figure \ref{fig4}(a)–(c) reveals that the $-k_y$ mode corresponds to a wave propagating in the negative refraction direction, while the $+k_y$ mode aligns with normal refraction. These results validate our hypothesis regarding the simultaneous occurrence of temporal and spatial reflections. Upon scattering at the interface, the two modes separate in $(\omega,k_y)$ space and subsequently propagate in distinct directions as they evolve in time.

\begin{figure*}[htp]
	\centering
	\subfigure[]{\includegraphics[width=0.30\linewidth]{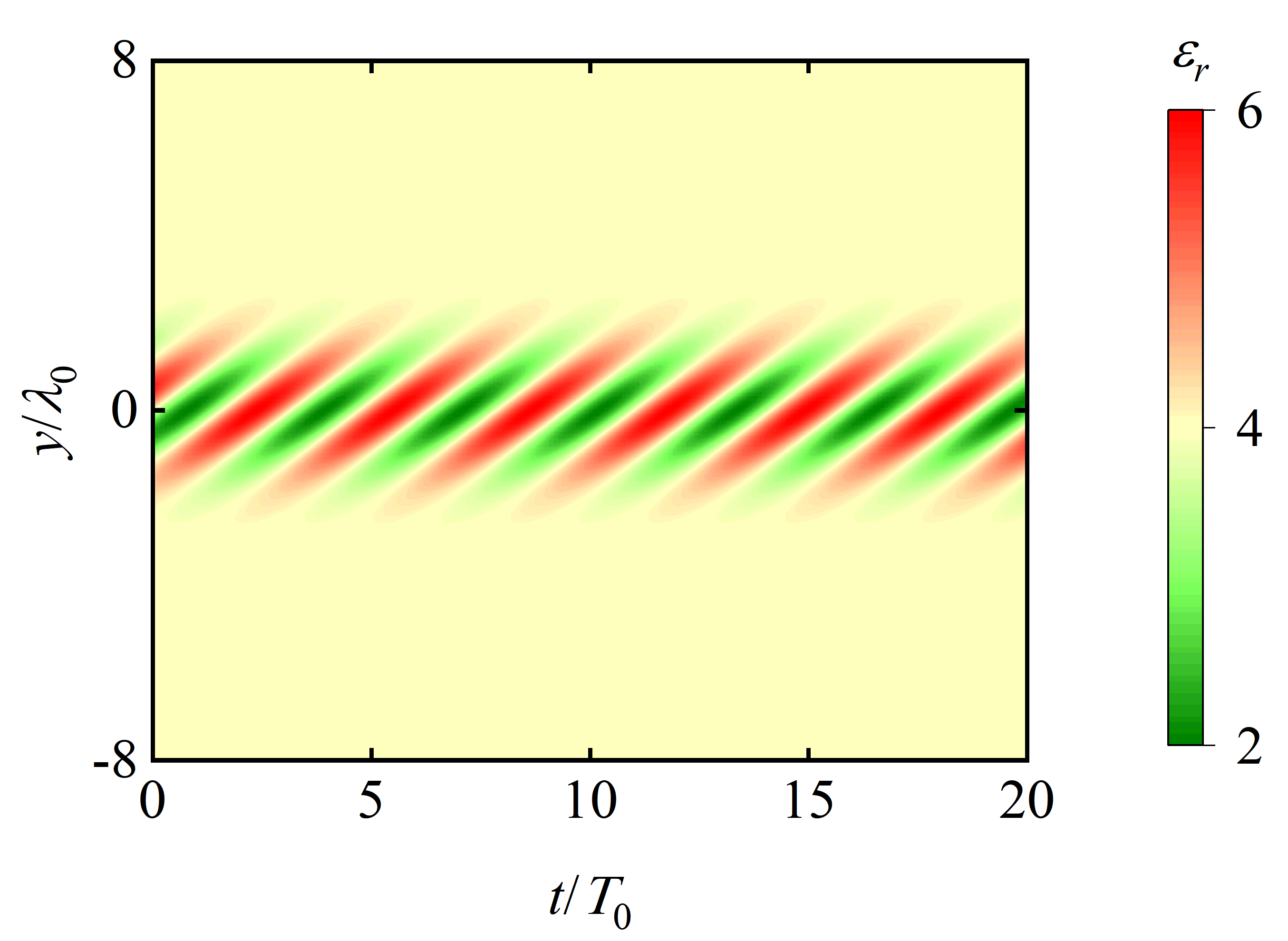}}
	\hspace{0.5em}
	\subfigure[]{\includegraphics[width=0.33\linewidth]{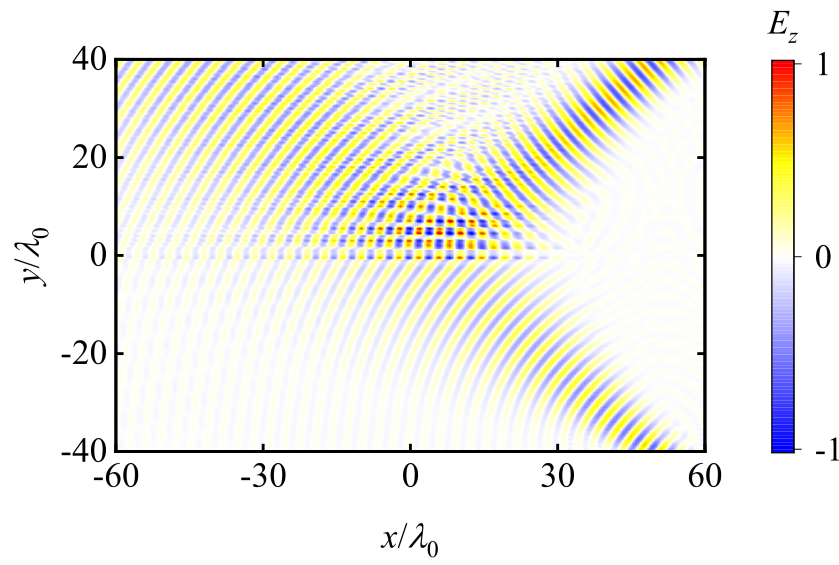}}
	\hspace{0.5em}
	\subfigure[]{\includegraphics[width=0.285\linewidth]{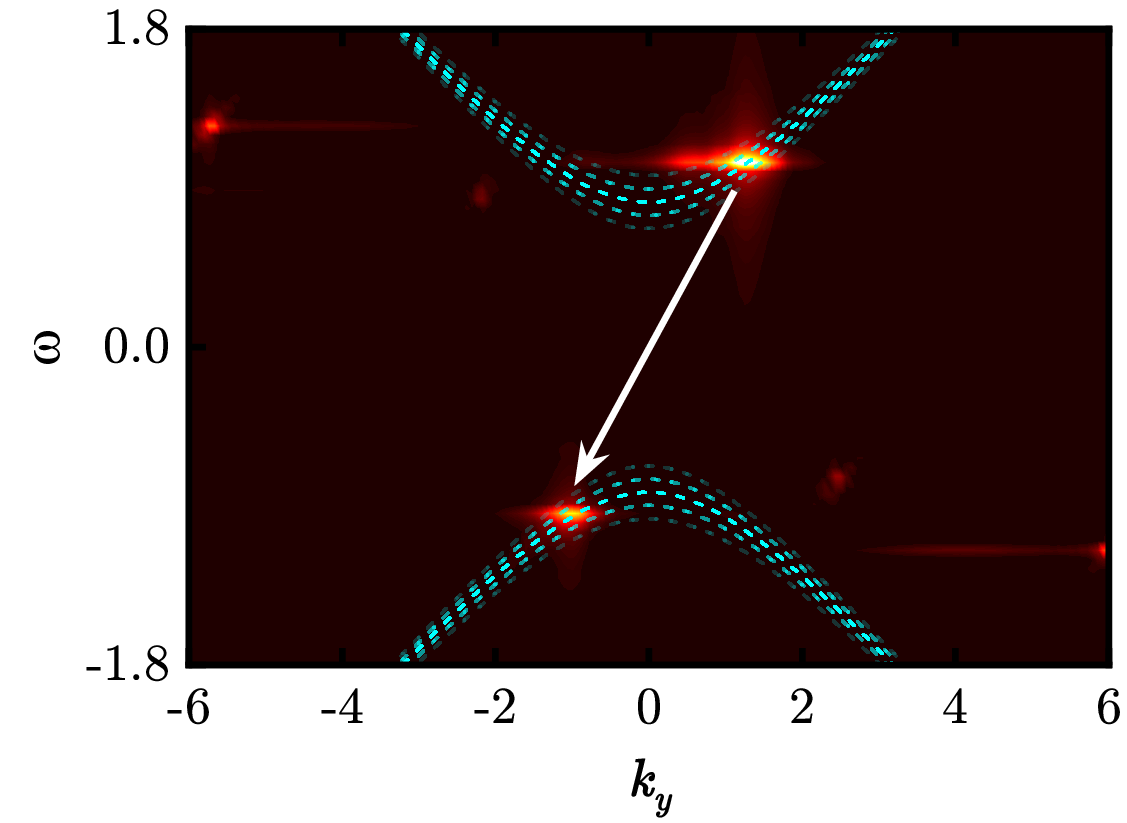}}
	\caption{
		Spatiotemporal flat lens proposal. (a) Spatiotemporal distribution of the permittivity $\varepsilon(t,y)$. (b) Spatial field distribution demonstrating the negative refraction phenomenon occurring within the modulated structure. (c) Spectral profile of the incident and scattered modes in $(\omega,k_y)$ space. Multiple dashed lines denote the $\omega-k_y$ dispersion relations for the background medium under oblique incidence, accounting for the finite spread in the tangential wave-vector component $kx$.}
	\label{fig5}
\end{figure*}

Since operational limitations such as infinite spatial extent and temporal non-repeatability constrain the practical application of moving interfaces, we propose a more realistic spatiotemporal optical device based on this principle. Specifically, we introduce a localized spatiotemporal permittivity modulation of the form:
$\Delta(y,t)=a\sin(\Delta k_y y -\Delta\omega t)\exp(-(y-y_0)^2/\Delta y^2)$
This configuration allows for a controlled shift in the normal wave vector $\Delta k_y$ and frequency $\Delta\omega$ while restricting the interaction to a spatial scale defined by $\Delta y$. Crucially, this localized modulation preserves translational symmetry along the interface direction. To ensure that spatiotemporal co-reflection is permitted, $\Delta k_y$ and $\Delta\omega$ must satisfy the phase-matching conditions established in the preceding sections. This mechanism enables the realization of a spatiotemporal flat lens.

An illustrative example is presented in Figure \ref{fig5}, with the spatiotemporal distribution of the permittivity shown in Figure \ref{fig5}(a). Here, $T_0=2\pi/\omega_0$. We set the background permittivity to 4 and the modulation amplitude to $a=2$. For an incident electromagnetic wave with a normalized frequency of $\omega_0$, the modulation parameters are set to $\Delta \omega=2 \omega_0$ and $\Delta k_y = 2.646 k_0$, with a spatial width of $\Delta y = 2\lambda_0$.

An obliquely incident Gaussian beam, defined by its central frequency $\omega=\omega_0$ and tangential wave vector $k_x= 1.5k_0$, is directed onto the modulated structure as shown in Figure \ref{fig5}(b). A pronounced negative refraction phenomenon is observed, with a near-total suppression of any equivalent reflective components. This behavior is further corroborated by the spectral analysis in Figure \ref{fig5}(c), which maps the corresponding dispersion relations and the distribution of incident and scattered modes in $(\omega,k_y)$ space. As predicted, the scattered wave emerges in the third quadrant, thereby implying the characteristic signatures of negative refraction.

\begin{figure}[htp]
	\centering
	\subfigure[]{\includegraphics[width=0.4\linewidth]{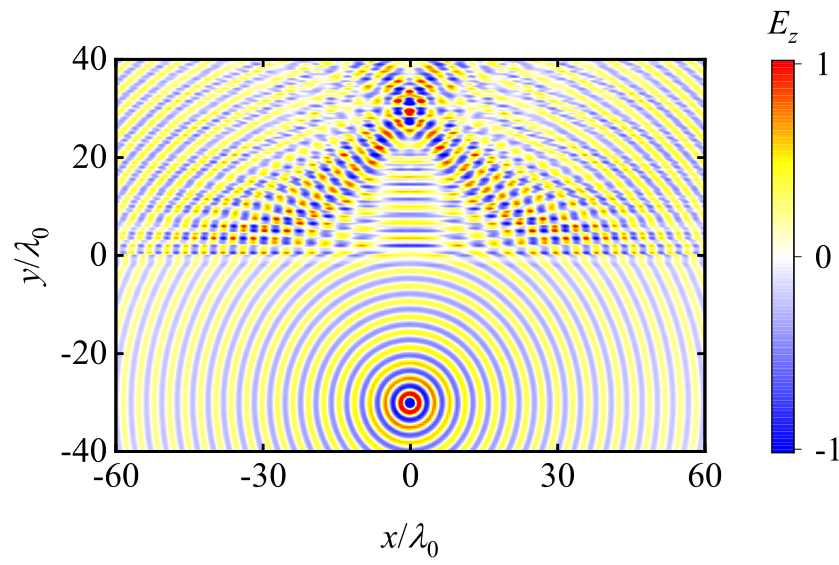}}
	\hspace{1em}
	\subfigure[]{\includegraphics[width=0.4\linewidth]{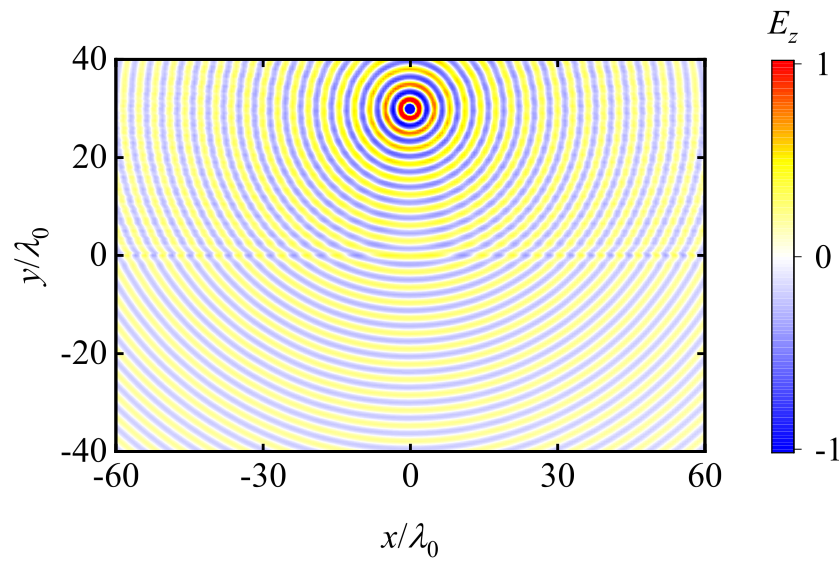}}

	\caption{Nonreciprocal wave focusing phenomenon. (a) A point source placed below the spatiotemporal lens; the negatively refracted wave converges at the symmetric position. (b) The point source is repositioned above the spatiotemporal lens. In this configuration, the interaction between the incident electromagnetic waves and the spatiotemporal modulation remains negligible, with no observable scattered components. }
	\label{fig6}
\end{figure}

It is well established that negative refraction enables the formation of a perfect unity-magnification real image. Our proposed architecture similarly exhibits a nonreciprocal imaging effect. As illustrated in Figure \ref{fig6}(a), a point source positioned below the spatiotemporal lens generates a scattered response for waves within a specific transverse wave-vector range. These components undergo negative refraction and converge at a symmetric focal point on the opposite side of the interface. Notably, the structure does not support the generation of significant reflective components.

Conversely, when the point source is placed above the spatiotemporal lens, as shown in Figure \ref{fig6}(b), the target frequency and wave vector produced by the modulation fail to satisfy the dispersion relations of the medium. Consequently, the interaction is nearly suppressed, and the generation of scattered waves is negligible.

It is noteworthy that this architecture possesses the predictable capability to propagate evanescent waves. By maintaining translational symmetry parallel to the interface, the structure preserves the transverse wave-vector components of the evanescent modes while coupling them into propagating states characterized by higher frequencies and positive normal wave vectors. However, a perfect lens remains unattainable, as the system responds only to modes within a specific range of transverse wave vectors. Nevertheless, the underlying mechanism differs fundamentally from traditional methods of achieving negative refraction in artificial media. By expanding the functional scope of spatiotemporal modulations, this work is expected to catalyze further research within the fields of time-varying photonics and sub-wavelength optics.

In summary, we have demonstrated that the simultaneous occurrence of temporal and spatial reflections at a moving interface results in an effective negative refraction regime. Notably, this phenomenon does not require the breaking of translational symmetry along the interface. While this spatiotemporal co-reflection is forbidden in one-dimensional media with linear dispersion, it becomes physically realizable in dispersive mediums, such as those characterized by a Drude model. Within a specific range of superluminal modulation velocities, this mechanism generates two distinct refractive modes, one of which corresponds to the spatiotemporal co-reflection case. Under oblique incidence, this mode appears as negative refraction through a concurrent sign reversal of both the frequency and the normal wave-vector component.

Through numerical simulations, we have validated the efficacy of this phenomenon and leveraged it to design a spatially limited spatiotemporal flat lens. This device facilitates the formation of a unity-magnification real image and exhibits inherent nonreciprocal focusing properties. Our findings provide a distinct physical route toward achieving negative refraction that bypasses the requirements of conventional artificial media. We anticipate that these results will catalyze further developments in spatiotemporal metamaterials and time-varying metasurfaces, paving the way for advanced applications in nonreciprocal and sub-wavelength photonics.



\medskip

\medskip
\textbf{Acknowledgements} \par 
This work is supported by the National Natural Science Foundation of China (NSFC, Contracts No.12575213, 12505229 and U2541210), Natural Science Foundation of Heilongjiang Province of China (No.YQ2024A008), the China Radio Wave Propagation Research Institute Stable Support for Scientific Research Funding under Project (A251200020) and National Key R \& D Program of China (No.2025YFF0512000).

\medskip

%
\bibliographystyle{MSP}
\bibliography{ref}

\end{document}